\documentclass[10pt,letterpaper]{article}
\usepackage[top=0.85in,left=2.75in,footskip=0.75in]{geometry}

\usepackage{amsmath,amssymb}

\usepackage{changepage}

\usepackage{textcomp,marvosym}

\usepackage{cite}

\usepackage{nameref,hyperref}
\usepackage{multirow}
\newcommand{\minitab}[2][l]{\begin{tabular}{#1}#2\end{tabular}}
\usepackage{tikz}
\usepackage[outline]{contour}
\contourlength{1.2pt}

\usepackage[right]{lineno}

\usepackage[nopatch=eqnum]{microtype}
\DisableLigatures[f]{encoding = *, family = * }

\usepackage{array}
\usepackage{multirow}
\usepackage{booktabs}

\usepackage[normalem]{ulem}

\def\R{\mathbb{R}}

\def\G{\mathcal{G}}

\def\D{\mathcal{D}}

\usepackage{color}
\newcommand\przemek[1]{{  #1}}

\newcolumntype{+}{!{\vrule width 2pt}}

\newlength\savedwidth

\raggedright
\usepackage{ragged2e}
\justifying

\usepackage[aboveskip=1pt,labelfont=bf,labelsep=period,justification=raggedright,singlelinecheck=off]{caption}

\makeatletter
\renewcommand{\@biblabel}[1]{\quad#1.}
\makeatother

\usepackage{lastpage,fancyhdr,graphicx}
\usepackage{epstopdf}
\fancyheadoffset[L]{2.25in}
\fancyfootoffset[L]{2.25in}
\def\our{MeVGAN}

\begin{document}
\vspace*{0.2in}

\begin{flushleft}
{\Large \our{}: GAN-based Plugin Model for Video Generation with Applications in Colonoscopy} 
\newline
\\
Ł. Struski\textsuperscript{1,3,*},
T. Urbańczyk\textsuperscript{2,3},
K. Bucki\textsuperscript{3},
B. Cupia\l{}\textsuperscript{1},
A. Kaczyńska\textsuperscript{1},
P. Spurek\textsuperscript{1,3},
J. Tabor\textsuperscript{1,3},
\\
\bigskip
\textbf{1} Jagiellonian University\\
\textbf{2} Marian Smoluchowski Institute of Physics, Jagiellonian University, S. Łojasiewicza 11,
30–348 Kraków, Poland\\
\textbf{3} Skopia Medical Center, J. Conrada 79, 31-357 Kraków, Poland\\

\bigskip

%
%





* lukasz.struski@uj.edu.pl

\end{flushleft}
\section*{Abstract}

Video generation is important, especially in medicine, as much data is given in this form. However,
video generation of high-resolution data is a very demanding task for generative models, due to the large need for memory. 
In this paper, we propose Memory Efficient Video GAN (\our{}) -- a Generative Adversarial Network (GAN)  which uses plugin-type architecture. We use a pre-trained 2D-image GAN and only add a simple neural network to construct respective trajectories in the noise space, so that the trajectory forwarded through the GAN model constructs a real-life video. We apply \our{} in the task of generating colonoscopy videos. Colonoscopy is an important medical procedure, especially beneficial in screening and managing colorectal cancer. However, because colonoscopy is difficult and time-consuming to learn, colonoscopy simulators are widely used in educating young colonoscopists.  
We show that \our{} can produce good quality synthetic colonoscopy videos, which can be potentially used in virtual simulators.


\section*{Introduction}

Video generation is an important field in AI, with many critical applications in biological domains and medicine \cite{golhar2022gan,wen2018colonoscopy}. However, video generation for medicine, where typically data has a high resolution, is very demanding for generative models due to the large need for memory.

\begin{figure}[!h]
    \begin{center}
    \begin{tikzpicture}[scale=1.0]
    \node[inner sep=0pt] (russell) at (0,0)
    {\includegraphics[width=0.7\textwidth]{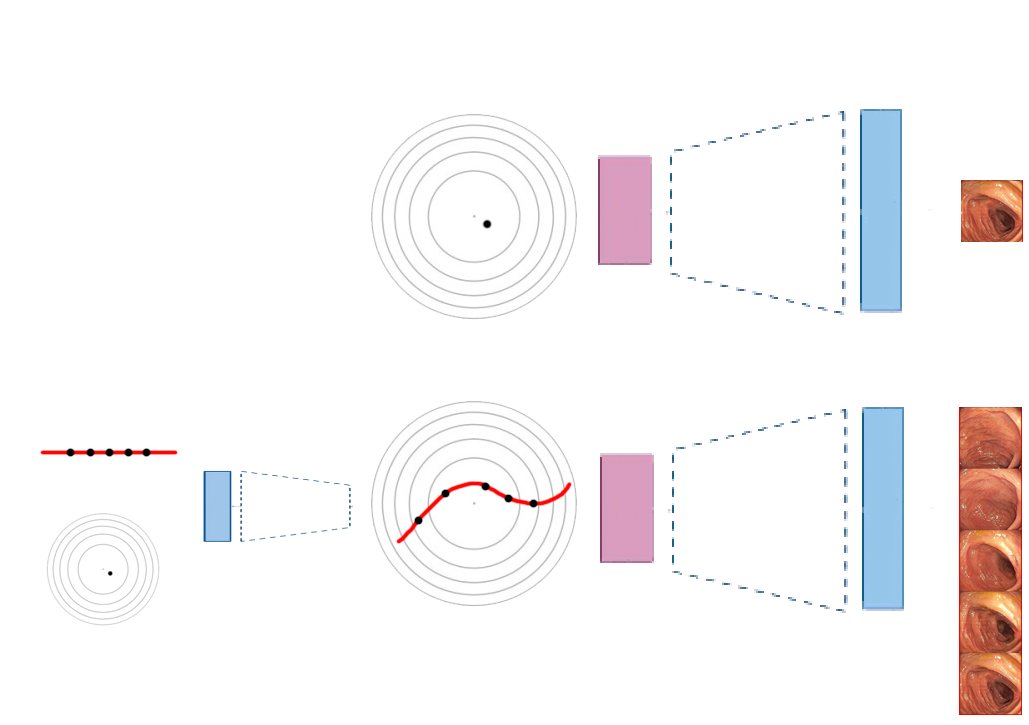} };
     \node[text width=3.5cm] at (1.4,1.5) {\footnotesize \contour{white}{  $ x $ }};
     \node[text width=3.5cm] at (3.25,1.4) {\footnotesize Generator};
     \node[text width=3.5cm] at (3.75,1.0) {\footnotesize $\G ( \cdot  )$};
     \node[text width=3.5cm] at (6.5,1.4) {\footnotesize $\G ( x )$};
     \node[text width=3.5cm] at (3.25,-1.2) {\footnotesize Generator};
     \node[text width=3.5cm] at (3.75,-1.6) {\footnotesize $\G ( \cdot  )$};
     \node[text width=3.5cm] at (6.5,-0.7) {\footnotesize $\G (\phi( t_1, z ))$};
     \node[text width=3.5cm] at (6.9,-1.7) { $ \vdots $};
     \node[text width=3.5cm] at (6.5,-3.1) {\footnotesize $\G (\phi( t_k, z ))$};
     \node[text width=3.5cm] at (-0.7,-1.3) {\footnotesize $\phi ( \cdot  )$};
     \node[text width=3.5cm] at (-1.9,-1.8) {\footnotesize $z$};
    \node[ text width=1.1cm] at (-3.75,-0.5) {\footnotesize \contour{white}{ $ t_1 \cdots t_k $ }};
    \node[text width=1.0cm] at (-1.2,-0.9) {\footnotesize \contour{white}{ $ \phi(t_1, z) \cdots \phi(t_k, z) $}};
     
    \node[text width=6.5cm] at (1.0,3.0) {Pre-trained classic 2D GAN generator};
    \node[text width=8.5cm] at (0.5,0.0) {\our{} with pre-trained classic 2D GAN generator};
    \end{tikzpicture}
    \caption{{\bf Top image:} Classic 2D GAN uses generator $\G$, which transforms samples from Gaussian distribution $x \sim N(0,I)$ into 2D images $\G(x)$ (see the top image). In practice, all separate frames are represented in Gaussian space. {\bf Bottom image:} \our{} model uses such a pre-trained model and incorporates an additional neural network $\phi(\cdot)$ which models the correct order of frames (see the red curve in the latent space). Since generator is pre-trained, we only need to train $\phi$.    } 
    \label{fig:architecture}
    \end{center}
\end{figure}
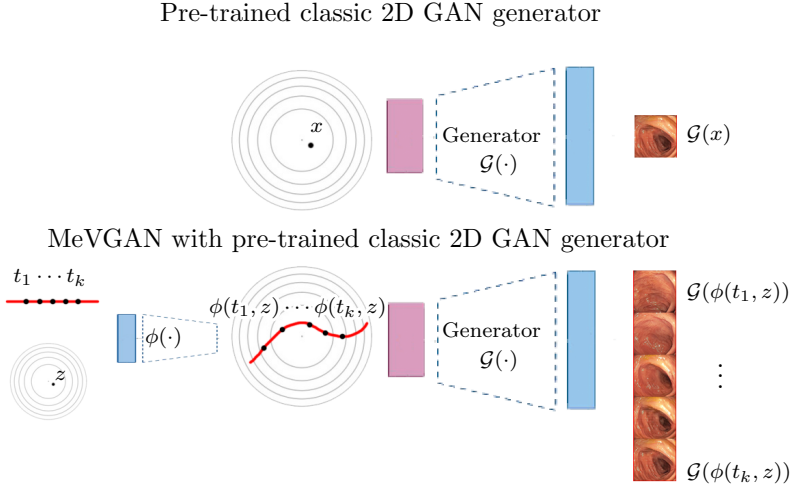

To generate high-quality images, we usually use GAN \cite{goodfellow2014generative} (Generative Adversarial Network), which uses a minimax game to model the data distribution.
GAN learns a generator network $\G$ that transforms samples from Gaussian noise $x \sim N(0, I)$ into an image $\G(x)$. The generator learns by playing against an adversarial discriminator network $\D$ aiming to distinguish between samples from the true data distribution and the generator's distribution. After training, we have GAN generator $\G(z)$; see the top model in Figure \ref{fig:architecture}.

We often need to produce many consecutive frames when we generate video. Therefore, the typical architecture that generates the video is large~\cite{acharya2018towards,clark2019adversarial}. Furthermore, the specificity of videos, a sequence of similar images,  forces the use of 3D convolution~\cite{acharya2018towards,clark2019adversarial,kahembwe2020lower,munoz2021temporal,saito2020train} or two discriminators~\cite{yu2021convolutional,tulyakov2018mocogan}.  
The biggest drawback of the existing methods is that we directly generate full movies. Such an approach requires high computational resources. \przemek{As a consequence, relatively short videos are produced. Alternatively, longer videos can be generated but with lower quality.}

To solve the above problem, we present \our{}\footnote{\url{https://github.com/gmum/MeVGAN}}, a method that uses a classic GAN model trained directly on 2D images that produces movies by creating the correct combination of relevant images. 
Conceptually \our{} can be seen as a plugin~\cite{wolczyk2022plugen} to GAN.
Thanks to such a solution, we can produce long and high-resolution videos.  
In \our{} we use frames from training videos to train the 2D GAN generator. Then, we train an additional neural network $\phi$, which learns to connect existing frames to the correct movies; see the bottom model in Figure~\ref{fig:architecture}. The neural network $\phi$ produces a curve in the latent space of the pre-trained 2D model to merge frames into the output video.  Such a solution generates high-quality videos.   

We test \our{} on the task of  generating high-quality videos on real colonoscopy images.
Colonoscopy is an important medical procedure that can be used for both diagnostic and therapeutic purposes. It is difficult and time-consuming to learn how to perform colonoscopy. Therefore, there is a problem with the procedure of training. Using live animals (usually pigs) for colonoscopy training can bring a lot of benefits \cite{wen2018colonoscopy}, but at the same time, it is complicated and may raise ethical concerns. To partially solve the above problems, various colonoscopy simulators have been created \cite{ GOODMAN20191}. There are many different approaches to construct such devices. Applying \our{} we can generate videos, which can be potentially applied to colonoscopy simulators. 

\przemek{
To summarize, the contributions of our work are the following:
\begin{itemize}
    \item We propose \our{}, a new memory efficient plugin-based generative model for videos.
    \item \our{} produces high-quality videos since we use pre-trained 2D-image GAN and only add neural network
to construct trajectories in the noise space to produce real-life video
    \item We show that \our{} can be used for modeling colonoscopy videos. 
\end{itemize}
}

\section*{Related works}

In this section, we present related works devoted to the three related scientific research directions. In the first one, we describe different 2D GAN models. In the second one, we present different approaches to video generation using GAN models. The third one is devoted to existing colonoscopy simulators.

\paragraph{Generative model} Generative modeling is a broad area of machine learning that deals with modeling a~joint data distribution. Roughly speaking,  generative models can produce examples similar to those already present in a training dataset $X$, but not exactly the~same. Generative models are one of the fastest-growing areas of deep learning. In recent years several generative models have been constructed. Among them Variational Autoencoders (VAE) \cite{kingma2014auto}, Wasserstein Autoencoders (WAE) \cite{tolstikhin2017wasserstein}, Generative Adversarial Networks (GAN) \cite{goodfellow2014generative}, Auto-regressive Models \cite{isola2017image} and~Flow-based Generative Models \cite{dinh2014nice,kingma2018glow}.

The quality of~generative image modeling has increased in recent years thanks to the GAN framework, which have solved many image generation tasks, like image-to-image translation\cite{isola2017image,zhu2017unpaired,taigman2016unsupervised,park2019semantic}, image super-resolution\cite{ledig2017photo,sonderby2016amortised}, and~text-to-image synthesis\cite{reed2016generative,hong2018inferring}.

GAN is a framework for training deep generative models using a minimax game. The goal is~to~learn a~generator distribution $P_{\G}(x)$ that matches the~real data distribution $P_{data}(x)$. GAN learns a~generator network $\G$ that generates samples from the~generator distribution $P_{\G}$ by transforming a~noise variable $z \sim P_{noise}(z)$ (usually Gaussian noise $N(0, I)$) into a~sample $\G(z)$. The generator learns by playing against an~adversarial discriminator network $\D$ which aims to~distinguish between samples from the~true data distribution $P_{data}$ and~the~generator's distribution $P_{\G}$. More formally, the~minimax game is~given by the~following expression:

$$
\min_{\G} \max_{\D} V(\D,\G) = 
\mathbb{E}_{x \sim P_{data}} [\log \D(x)] +  \mathbb{E}_{x \sim noise} [\log (1-\D(\G(x)))].
$$

The main advantage of GANs over other models is producing sharp images that are indistinguishable from the real ones. 
GANs are impressive regarding the visual quality of images sampled from the model, but the training process is often challenging and unstable.

In recent years, many researchers focused on~modifying the~vanilla GAN procedure to~improve stability of~the training process, by change of the~objective function to~Wasserstein distance (WGAN) \cite{arjovsky2017wasserstein}, restrictions on~the~gradient penalties \cite{gulrajani2017improved,kodali2017convergence},  Spectral Normalization \cite{miyato2018spectral}, imbalanced learning rates for generator and~discriminator\cite{gulrajani2017improved,miyato2018spectral},  Self-Attention mechanisms SAGAN~\cite{zhang2018self} and~progressively growing architectures such as ProGAN~\cite{karras2017progressive} 
 or StyleGAN~\cite{karras2019style}.

In addition to works improving the training stability, several modifications of the vanilla GAN architecture are dedicated to specific tasks, like generating textures \cite{hedjazi2021efficient}, producing images with different resolutions and~training on~a~single image \cite{shaham2019singan}.
Such methods enable GANs training on images with varying resolutions.

\paragraph{GAN for video}

Video GANs deal with multiple images with an additional time dimension. To solve such a problem, video GANs use many different strategies \cite{aldausari2022video}.

Video-GAN (VGAN) \cite{vondrick2016generating} is one of the first applications of GAN for video generating. The generator consists of two convolutional networks. The first is the 2D convolutional model for the static background, while the second is a 3D convolutional network that models moving objects in the foreground. In (FTGAN), authors add progressive architecture to model the the motion of an object is more effectively. 
MoCoGAN \cite{tulyakov2018mocogan} traverses $N$ latent points, one per frame, using recurrent neural networks RNNs. Like MoCoGAN, Temporal Generative Adversarial Nets (TGAN) \cite{saito2017temporal,saito2020train} use $N$ latent vectors for $N$ frames. However, each frame is generated from a latent vector in TGAN whereas in MoCoGAN, a frame is generated from a combination of a motion vector and a fixed content vector shared across the frames.
 Similarly, G3AN~\cite{wang2020g3an} proposes a three-stream generator to disentangle motion and appearance with a self-attention module.

\przemek{In \cite{natarajan2022dynamic} authors use skeletal pose information and person images as input and produce high-quality videos. In the generator phase, the proposed model uses a U-Net-like network
to generate target frames from skeletal poses.

In \cite{natarajan2022development} authors propose an end-to-end deep learning framework for sign language recognition, translation, and video generation.

}

Another approach is to use 3D convolutional networks instead of 2D convolutions \cite{acharya2018towards,clark2019adversarial,kahembwe2020lower,munoz2021temporal,saito2020train}. Dual video discriminator GAN (DVD-GAN) \cite{yu2021convolutional} applied BigGAN architecture to video generation. Similar to MoCoGAN, there are two discriminators to deal with the temporal and spatial aspects of a
video.

In \cite{skorokhodov2022stylegan}, authors build the model on top of StyleGAN2 \cite{karras2020training} and redesign its generator and discriminator networks for video
synthesis. In \cite{yu2022generating} authors use an implicit representation of video.
Some recent works also consider high-resolution video
synthesis \cite{fox2021stylevideogan}, but only with training in the latent space
of a pre-trained image generator.

Another approach uses a two-stream architecture for modeling different aspects of video: motion and content \cite{sun2020twostreamvan}. 
EncGAN3 \cite{yang2022encoder} also decomposes the video into two
streams representing content and movement but consists of three processing modules,
representing Encoder, Generator, and Discriminator, each trained separately.

\paragraph{Colonoscopy simulators}

Colonoscopy is an important medical procedure that can be used for both diagnostic and therapeutic purposes. It plays a very important role in the diagnosis and prevention of colorectal cancer (CRC), because it enables early detection and extraction of polyps, which are often the first stage of colorectal cancer which is one of the most prevalent and significant causes of morbidity and mortality in the developed world \cite{wen2018colonoscopy}.  Moreover, colonoscopy is also helpful in diagnosing many other diseases, such as ulcerative colitis, Crohn's disease, and diverticulosis \cite{KimYongIBD}.

Learning how to perform colonoscopy is difficult and time-consuming. Additionally, choosing the right training procedure is problematic. Statistical studies suggest that up to 700 performed procedures are required to gain proficiency \cite{patwardhan2016fellowship}.  Proper preparation of the doctor performing colonoscopy is very important from the point of view of the effectiveness of colorectal cancer prevention, as one of the main reasons for overlooking polyps during the examination is the inexperience of the endoscopist \cite{Papanikolaou}.
\przemek{It is worth mentioning that works are currently being carried out to automate the process of detecting of polyps in colonoscopy videos \cite{polypDetection2D3DCNN} which can significantly reduce the number of missed polyps. Similar automated systems have also been developed for other types of medical data, for example there exists frameworks for cervical cancer detection and classification\cite{elakkiya2021cervical,elakkiya2022imaging}.}

On the other hand, it is ethically questionable to train colonoscopy practitioners on real patients, as a poorly performed colonoscopy can have severe complications, including perforation, bacteriaemia, and hemorrhage \cite{fisher2011complications,latos2022colonoscopy}. Using live animals (usually pigs) for colonoscopy training can bring a lot of benefits\cite{wen2018colonoscopy}, but at the same time, it is complicated and may raise ethical concerns. To partially solve the above problems, different colonoscopy simulators have been created \cite{ GOODMAN20191}. These simulators use a variety of techniques, ranging from simple mechanical models \cite{Classen}, through composite devices that use explanted animal organs \cite{HOCHBERGER2005204}, to computerized virtual simulators which incorporate visual interface with haptics \cite{Koch2008, wen2018colonoscopy}. There are many different approaches for constructing such virtual simulator.  This paper uses a neural network approach to generate artificial videos of colonoscopy procedures. 

\przemek{The use of virtual simulators – which could be developed e.g. using video sequences created by MeVGAN – to train colonoscopists has numerous advantages. One of them is their high realism, which can be even greater than in the case of training on live animals. Thanks to this, it is possible to reduce the suffering of animals associated with using them for colonoscopic training. Another big advantage of virtual simulators is the ability to easily and effectively simulate complex medical procedures or disease cases (e.g. polyps or colorectal cancer), which are relatively rare. In our work, we have shown that the MeVGAN model can be used to generate video sequences where polyps are visible.  An important advantage of generative models (including MeVGAN) used to generate textures or video sequences for use in virtual simulators is their ability to anonymize sensitive medical data. Although data from real patients are used to train the model, it is virtually impossible to link the results returned by the model to data from a specific patient. According to \cite{synteticDataInHealthcare}, the use of synthetic data is a method to share medical datasets with a wider audience. Therefore,  using generative models can be beneficial from the point of view of protecting sensitive medical data of patients.}



\section*{Description of \our{}}

In this section, we describe our model. In \our{}, we use a pre-trained GAN model dedicated to 2D images and add neural network to adopt such a model for video generation. \przemek{In presented model we use ProGAN~\cite{karras2017progressive} as the backbone, therefore we first describe the classic ProGAN model for 2D images and  then we introduce \our{}.}

\paragraph{ProGAN} ProGAN~\cite{karras2017progressive} is a classic GAN with a minimax game. It consists of a generator network $\G_{ProGAN}$ that transfers samples from the~prior Gaussian noise $N(0, I)$ into 2D images, and  the  discriminator network $\D_{ProGAN}$ that aims to~distinguish between samples from the true and learned data distribution.
The main advantage of the ProGAN model is its architecture. The model starts with low-resolution images and then progressively increase the resolution by adding layers to the generator and the discriminator.  This incremental nature allows the training first to discover the large-scale structure of the image distribution and then shift attention to increasingly finer-scale details instead of having to learn full scales simultaneously.

Progressive training has several benefits. The train procedure on smaller images is substantially more stable because there is less class information and fewer modes.  Another benefit is the reduced training time. With progressively growing GANs, most iterations are done at lower resolutions.

ProGAN can be easily trained in colonoscopy images with arbitrary resolution. \przemek{In \our{} we used the PyTorch implementation of ProGAN\footnote{\url{https://github.com/facebookresearch/pytorch_GAN_zoo}}, trained on colonoscopy images up to resolution of 1024px. We mostly used default training hyperparameters, which includes using WGAN-GP loss \cite{gulrajani2017improved}. The only modification we did was setting the batch size to 8 for resolution up to 256px, and then decreasing it to 4 for the rest of the training.}

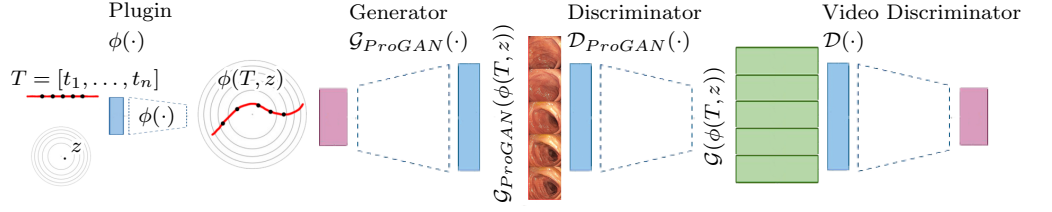
\begin{figure}[!h]
    \begin{center}
    \begin{tikzpicture}[scale=1.0]
    \node[inner sep=0pt] (russell) at (0,0)
    {\includegraphics[width=1.0\textwidth]{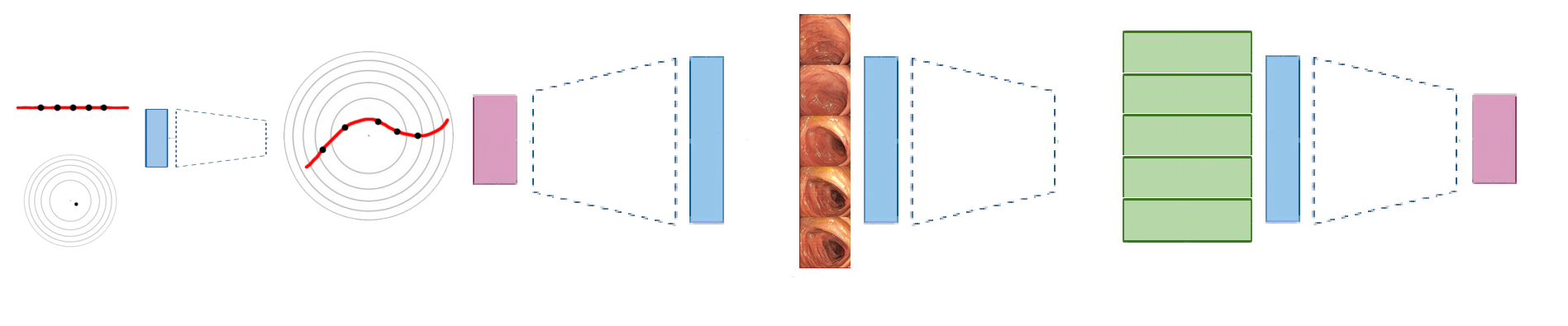} };
     \node[text width=3.5cm] at (-3.7,1.6) {\footnotesize Plugin};
     \node[text width=3.5cm] at (-3.7,1.2) {\footnotesize $\phi ( \cdot  )$};
     \node[text width=3.5cm] at (-0.5,1.6) {\footnotesize Generator};
     \node[text width=3.5cm] at (-0.5,1.2) {\footnotesize $\G_{ProGAN} ( \cdot  )$};
     \node[text width=3.5cm] at (2.4,1.6) {\footnotesize Discriminator};
     \node[text width=3.5cm] at (2.4,1.2) {\footnotesize $\D_{ProGAN} ( \cdot  )$};
     \node[text width=3.5cm] at (5.8,1.6) {\footnotesize Video Discriminator};
     \node[text width=3.5cm] at (5.8,1.2) {\footnotesize $\D ( \cdot  )$};

\node[text width=3.5cm, rotate=90,] at (-0.2,0.8) {\footnotesize $\G_{ProGAN}(\phi(T,z))$};

\node[text width=3.5cm, rotate=90,] at (2.6, 1.3) {\footnotesize $\G(\phi(T,z))$};
     
     \node[text width=3.5cm] at (-3.3,0.2) {\footnotesize $\phi ( \cdot  )$};
    \node[ text width=1.1cm] at (-6.3,0.7) {\footnotesize \contour{white}{ $ T = [t_1, \ldots, t_n] $ }};
    \node[ text width=1.1cm] at (-5.5,-0.2) {\footnotesize \contour{white}{ $ z $ }};
    \node[text width=1.0cm] at (-3.6,0.7) {\footnotesize \contour{white}{ $ \phi(T, z)  $}};
     
    \end{tikzpicture}
    \caption{\our{} model uses a pre-trained ProGAN model, which consists of $\G_{ProGAN}$ and $\D_{ProGAN}$. In practice, all separate frames are represented in ProGAN's Gaussian space. \our{} model uses such a pre-trained model and incorporates additional neural networks: plugin $\phi$ and video discriminator $\D$, which are responsible for the correct sequence of frames. $ \phi$ transfers Gaussian noise $z$ and time indexes $t_1, \ldots, t_n$ into ProGAN's latent codes for separate frames. Then we use pre-trained ProGAN generator $\G_{ProGAN}$ to obtain video frames $\G_{ProGAN}(\phi(T,z))$. Before we use the video discriminator, we transfer frames by pre-trained ProGAN discriminator (without the last layer) to obtain a low-dimensional representation of movies $\G(\phi(T,z))$. Such full video representation goes to classic 2D video discriminator $\D$.   } 
    \label{fig:full_architecture}
    \end{center}
\end{figure}

\paragraph{\our{} } In this part of the section, we present our extension of a generative model that was originally trained on 2D data (images). We assume that we have the pre-trained ProGAN model on frames from training videos so we have generator $\G_{ProGAN}$ and discriminator $\D_{ProGAN}$ dedicated for 2D images. In \our{}, these networks will be frozen. We aim to train two additional networks: plugin $\phi$ and video discriminator $D$. The first neural network allows the ProGAN generator $\G_{ProGAN}$ to model the subsequent frames of generated video. On the input of plugin network $\phi$, we have a Gaussian noise $z$ and timeline (consecutive indexes of frames in the video) 
$$
T = [t_1, \ldots, t_n].
$$ 
The Plugin $\phi$ transfers such representation to obtain the latent codes of video frames 
$$ 
\phi(T,z) = [\phi(t_1,z), \ldots, \phi(t_n,z)].
$$
Plugin consists of three fully connected layers that uses the temporal information (i.e., the order of frames in the video sequence) to produce a sequence of $n$ noise vectors with a $(n, 512)$ shape.

The ProGAN generator $\G_{ProGAN}$ transfers this sequence of noise vectors into video frames 
$$
\G_{ProGAN}(\phi(T,z)) = [\G_{ProGAN}(\phi(t_1,z)), \ldots, \G_{ProGAN}(\phi(t_n,z))].
$$ 
Subsequently, the discriminator is used to enforce the smooth transition between consecutive frames in the output video. In the majority of existing solutions discriminators utilize 3D convolutional layers, however in \our{}, we operate on the low-dimensional representation of images. In practice, we use pre-trained ProGAN discriminator $\D_{ProGAN}$ without the last layer to extract features. Such representations are combined into a single tensor
$$
\G_{\phi}( T, z ) = 
[ \D_{ProGAN}(\G_{ProGAN}(\phi(t_1,z))), \ldots, \D_{ProGAN}(\G_{ProGAN}(\phi(t_n,z))) ].
$$
In consequence, we obtain \our{} generator $\G_{\phi}$, which consist of three neural networks: plugin $\phi$,  pre-trained ProGAN generator $\G_{ProGAN}$ and pre-trained ProGAN discriminator $\D_{ProGAN}$.
Using the discriminator $\D_{ProGAN}$ as a feature extractor, we can use classic 2D video discriminator $\D$ instead of 3D convolutions. 

This approach enables the discriminator to output a single value for a sequence of images, similar to a traditional discriminator. Treating the sequence of features as an image makes it easier for the discriminator to differentiate between a real movie and a sequence of unordered frames. Our discriminator architecture offers several advantages, including improved computational efficiency. By utilizing the ProGAN architecture, we can also ensure that our discriminator is well-suited for use with the generator in our proposed extension.

Our model is trained analogically to classic GAN by minimizing minimax game 
$$
\min_{\phi} \max_{\D}  
\mathbb{E}_{x \sim P_{data}} [\log \D(x)] +  \mathbb{E}_{z \sim noise} [\log (1-\D( \D_{ProGAN}( \G_{\phi}( t_1, \ldots, z_{n}, x ) ))].
$$

\begin{table}[h!]
\begin{center}
\przemek{
\begin{tabular}{ |c|c|c| } 
    \hline
    
    \multicolumn{2}{|c|}{Plugin $\phi$} & Video Discriminator $\D$  \\ 
    \hline\hline

    $z\in \R^{8 \times 2047}$ & $T \in \mathbb{R}^{8}$ & $\G(\phi(T,z))\in \mathbb{R}^{1 \times 1 \times 8 \times 512}$ \\ 
    \hline

    \multicolumn{2}{|c|}{Concat} & \multirow{3}{*}{\minitab[c]{Conv2D(1, 16, (3, 10), (1, 2)) \\ ReLU}}\\
    \multicolumn{2}{|c|}{Linear(2048, 1535)} &  \\
    \multicolumn{2}{|c|}{ReLU} & \\
    \hline

    \multicolumn{2}{|c|}{Concat} & \multirow{3}{*}{\minitab[c]{Conv2D(16, 8, (3, 8), (1, 2)) \\ ReLU}}\\
    \multicolumn{2}{|c|}{Linear(1536, 1023)} &  \\
    \multicolumn{2}{|c|}{ReLU}                &  \\
    \hline
    
    \multicolumn{2}{|c|}{Concat} & \multirow{3}{*}{\minitab[c]{Conv2D(8, 4, (3, 6), (1, 2)) \\ ReLU}}\\
    \multicolumn{2}{|c|}{Linear(1024, 511)}  &  \\
    \multicolumn{2}{|c|}{ReLU}                &  \\
    \hline
    
    \multicolumn{2}{|c|}{Concat} & \multirow{3}{*}{\minitab[c]{Conv2D(4, 1, (2, 6), (1, 2)) \\ ReLU}}\\
    \multicolumn{2}{|c|}{Linear(512, 512)}   &  \\ 
    \multicolumn{2}{|c|}{-}               &  \\ 
    \hline

    \multicolumn{2}{|c|}{\multirow{2}{*}{Norm}} & Linear(27, 1) \\
    \multicolumn{2}{|c|}{} & Sigmoid \\
    \hline

\end{tabular}
}
\end{center}
\caption{
\przemek{Detailed description of Plugin $\phi$ and Video Discriminator $\D$ architecture. \textbf{Plugin $\phi$} takes random noise vector $z$ and temporal vector $T$ as an input. At each of the four layers, it first concatenates temporal vector $T$ with previous layer output, then forwards it through the Linear layer. The first three layers  additionally use ReLU activation. After the fourth layer, the output vector
is normalized to lie on a hypersphere. \textbf{Video Discriminator $\D$} takes low-dimensional representation of generated video $\G(\phi(T,z))$ as input. At each of the four layers, it applies 2D Convolution and ReLU activation. At the end, the output is flattened and forwarded through the Linear layer and Sigmoid function, to obtain probability. Conv2D parameters describe, as follows, the number of input and output channels, kernel size and stride. Linear layer parameters refers to the number of input and output channels.}}
\label{tab:arch_details}
\end{table}

\przemek{In each step the video discriminator is taught to distinguish between real and fake videos, while the plugin network is taught to fool the discriminator, by generating (together with ProGAN's generator) possibly the most realistic videos. Both networks are trained using Adam optimizer with learning rate of 0.0002, $\beta_{1}=0.5$, $\beta_{2}=0.999$ and Binary Cross Entropy loss, for 50 epochs on NVIDIA Tesla V100 SXM2 32GB GPU.}




\przemek{
The MeVGAN model stands out as an innovative approach, replete with both limitations and distinct advantages vis-à-vis its competitor, the Temporal GAN model. A central constraint of the MeVGAN model lies in its profound reliance on a pre-trained generative model. The quality of the generated images is intricately intertwined with the performance and robustness of this foundational model. Consequently, any limitations or biases inherent in the pre-trained model may seamlessly permeate into the generated content.

Diverging from the Temporal GAN model, which commences training from scratch for the entire architecture, MeVGAN adopts a different approach. It incorporates a generator that remains fixed after pre-training. This static characteristic can obstruct the model's adaptability and its capacity for fine-tuning, particularly in response to evolving data distributions over time.

However, this unique structural aspect of our model unveils a multitude of advantages. Foremost among these is its ease of learning. By harnessing a pre-trained generative model, MeVGAN circumvents the formidable challenge of learning a generative model from the beginning. This pragmatic choice often results in faster convergence during training. Moreover, MeVGAN's reliance on a pre-trained generator typically entails fewer parameters compared to models that commence training anew for the entire architecture. This streamlined parameterization can significantly reduce computational demands and resource consumption, rendering MeVGAN a more efficient choice for various use cases.

MeVGAN's core focus revolves around the task of learning noise paths to generate videos. This task, in contrast to training a model alongside a generator for individual frames, is notably more manageable. Furthermore, when the noise space exhibits 'richness', such a learning paradigm can yield video content that is remarkably consistent and aesthetically pleasing, especially when contrasted with models that fail to capture such intricate noise patterns effectively.
}

\section*{Experiments}

The experiments section consists of two main parts. In the first subsection, we compare our method with the baseline approach, Temporal GAN v2~\cite{saito2020train}, on classic benchmarks. The second subsection shows how our model works on colonoscopy videos. 

\subsection*{Comparison with baseline model}

This section compares our method with the baseline approach, Temporal GAN v2~\cite{saito2020train}. For this purpose, we compared the performance of these methods using practical datasets such as UCF-101~\cite{soomro2012ucf101}.
It is a widely used benchmark dataset for action recognition in videos, consisting of 101 different action categories, each containing at least 100 videos. The average length of a video clip is 6 seconds. The videos in UCF-101 cover various actions, including sports activities, human-object interactions, and animal behavior. The dataset provides a challenging benchmark for action recognition algorithms, as the videos contain a lot of variability in viewpoint, lighting conditions, background clutter, and actor appearance. For our experiments, we selected four categories: BalanceBeam, BaseballPitch, Skiing, and TaiChi. 

\begin{figure}
    \centering
    \renewcommand{\arraystretch}{0.1}
    \begin{tabular}{ @{}r@{\;}l@{} }
    \multirow{3}{*}{\rotatebox{90}{\makebox[0pt][c]{BalanceBeam}}} &
    \includegraphics[width=.95\textwidth]{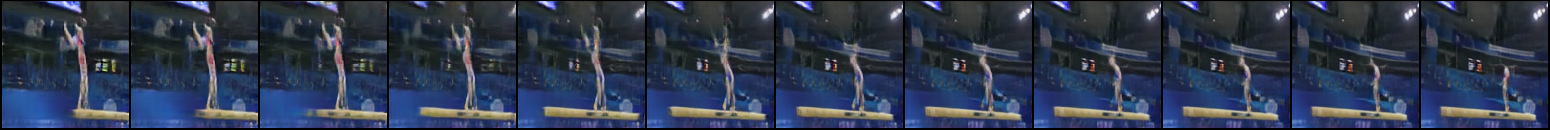} \\
    & \includegraphics[width=.95\textwidth]{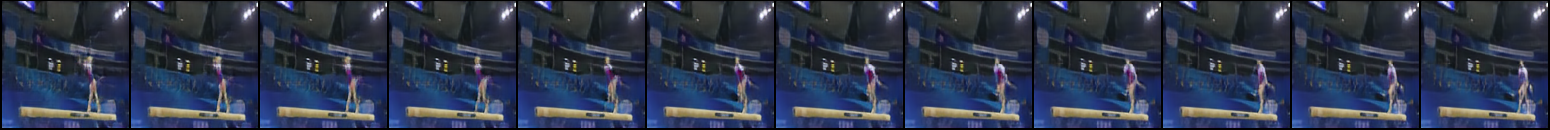} \\
    & \includegraphics[width=.95\textwidth]{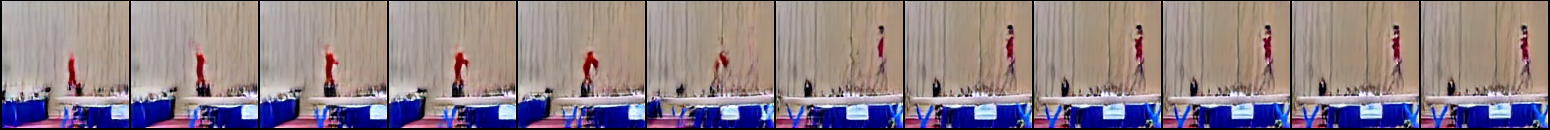} \\[2pt]

    \multirow{3}{*}{\rotatebox{90}{\makebox[0pt][c]{BaseballPitch}}} &
    \includegraphics[width=.95\textwidth]{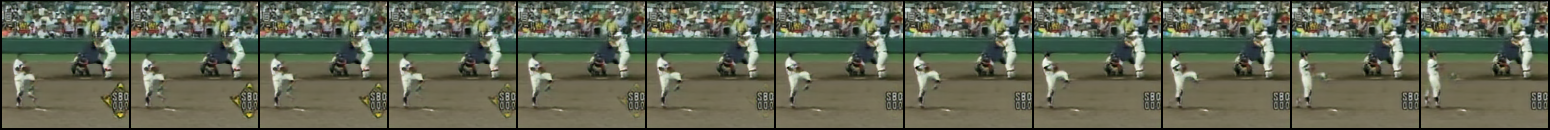} \\
    & \includegraphics[width=.95\textwidth]{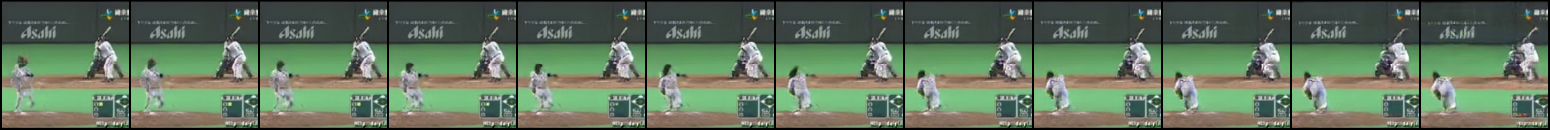} \\
    & \includegraphics[width=.95\textwidth]{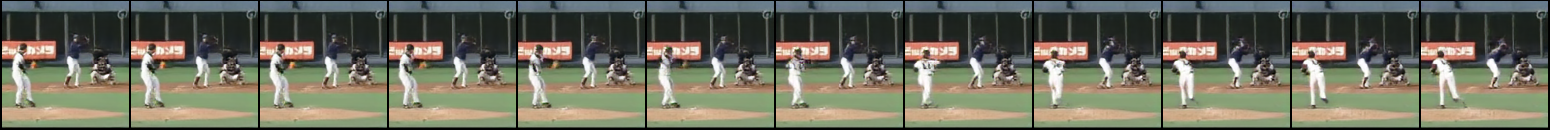} \\[2pt]

    \multirow{3}{*}{\rotatebox{90}{\makebox[0pt][c]{Skiing}}} &
    \includegraphics[width=.95\textwidth]{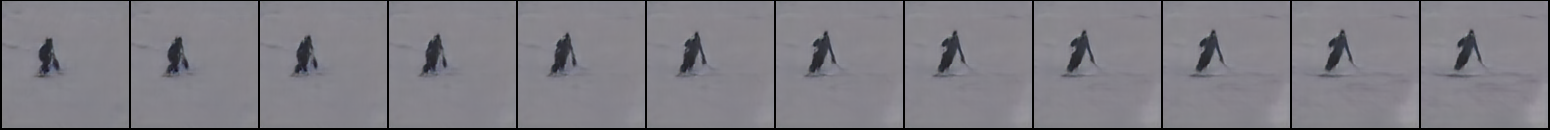} \\
    & \includegraphics[width=.95\textwidth]{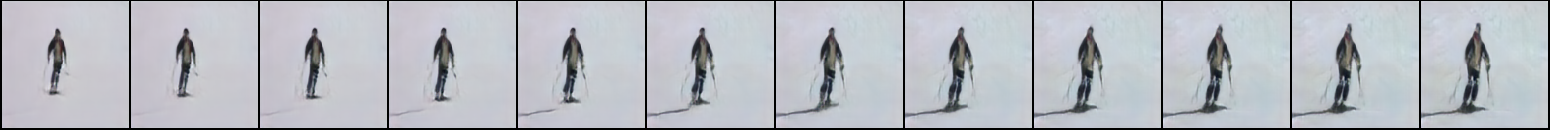} \\
    & \includegraphics[width=.95\textwidth]{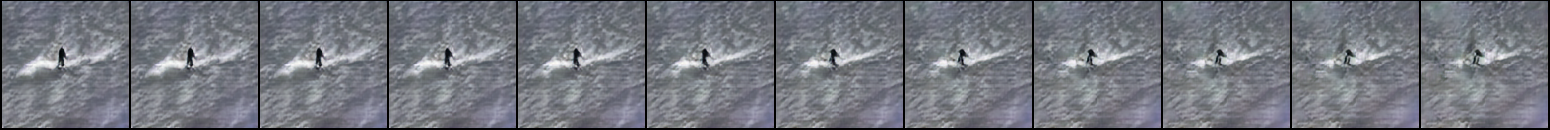} \\[2pt]

    \multirow{3}{*}{\rotatebox{90}{\makebox[0pt][c]{TaiChi}}} &
    \includegraphics[width=.95\textwidth]{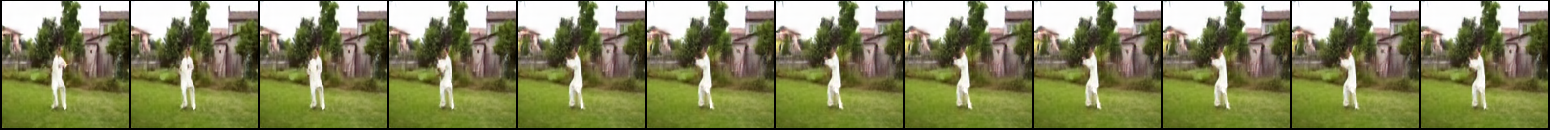} \\
    & \includegraphics[width=.95\textwidth]{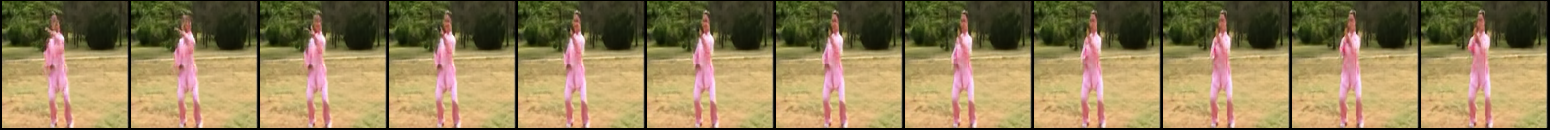} \\
    & \includegraphics[width=.95\textwidth]{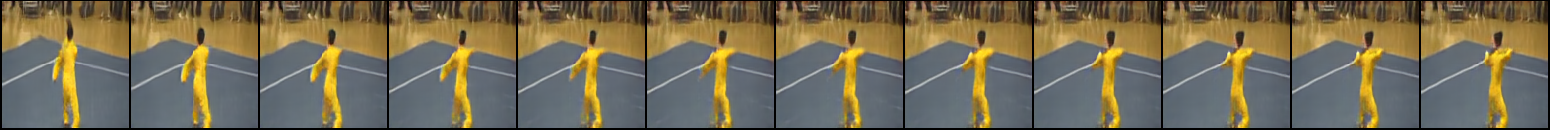} \\
    \end{tabular}
    \caption{\przemek{An example of video frames generated be MeVGAN model presented in this work, which was trained on four categories of the UCF-101 dataset. Each row presents twelve generated consecutive frames from the movie belonging to category described in the right part of the Figure. Compare to Fig. \ref{fig:our_generated_TG}, which presents an example of video frames generated by TGANv2.}}
    \label{fig:our_generated}
\end{figure}

\paragraph{Setup} For each selected dataset category, we trained the ProGAN model to a resolution of $128\times128$ pixels. We expanded the initial part of the ProGAN model in such a way that it generated sequences of n noises $(n, 512)$ for the ProGAN generator, from a single noise vector of size 2048. By using the pre-trained ProGAN, we could focus solely on training the initial component responsible for generating sequences of noise vectors that would result in smooth and continuous video sequences. Otherwise, we would have to start from scratch and train the model to generate entire videos from data. Figure~\ref{fig:our_generated} provides several examples of video clips generated by our model. 

\begin{figure}
    \centering
    \renewcommand{\arraystretch}{0.1}
    \begin{tabular}{ @{}r@{\;}l@{} }
    \multirow{3}{*}{\rotatebox{90}{\makebox[0pt][c]{BalanceBeam}}} &
    \includegraphics[width=.95\textwidth]{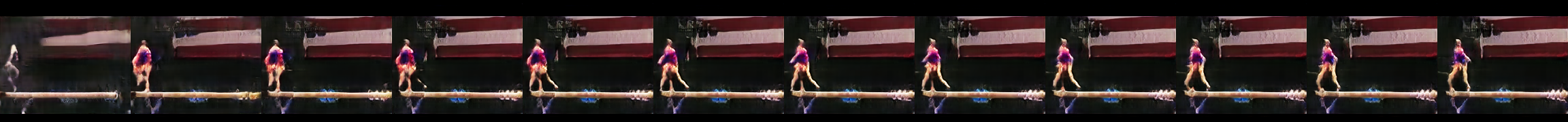} \\
    & \includegraphics[width=.95\textwidth]{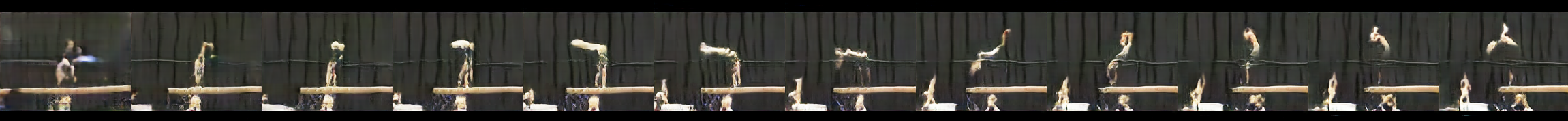} \\
    & \includegraphics[width=.95\textwidth]{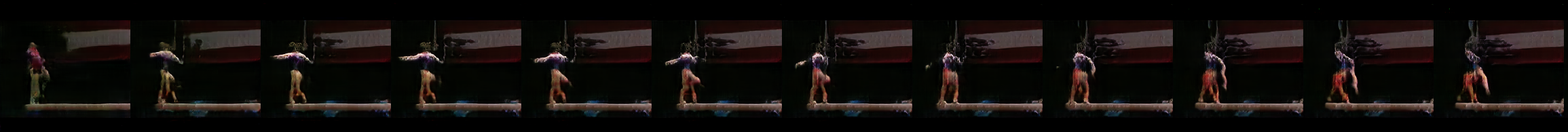} \\[2pt]

    \multirow{3}{*}{\rotatebox{90}{\makebox[0pt][c]{BaseballPitch}}} &
    \includegraphics[width=.95\textwidth]{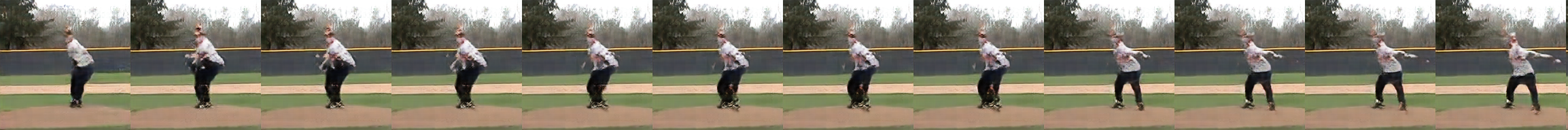} \\
    & \includegraphics[width=.95\textwidth]{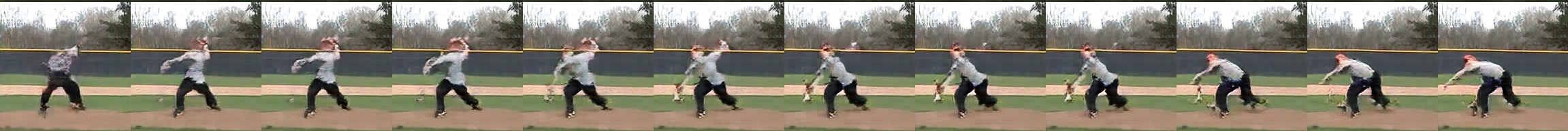} \\
    & \includegraphics[width=.95\textwidth]{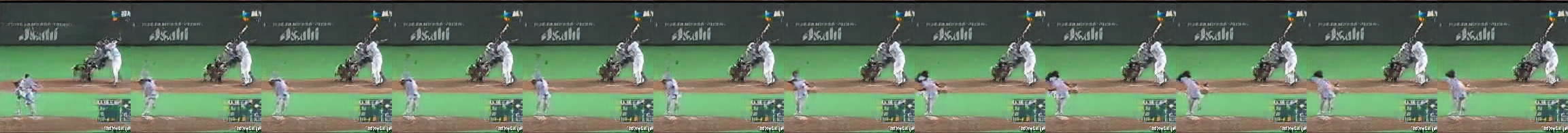} \\[2pt]

    \multirow{3}{*}{\rotatebox{90}{\makebox[0pt][c]{Skiing}}} &
    \includegraphics[width=.95\textwidth]{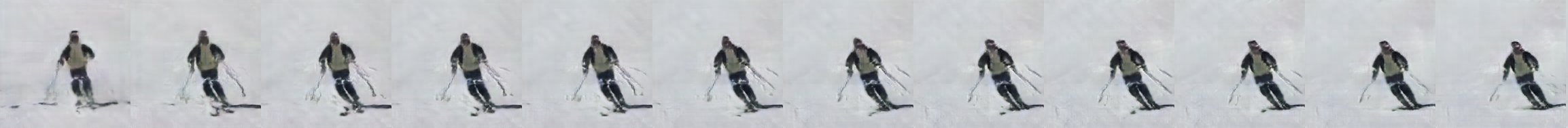} \\
    & \includegraphics[width=.95\textwidth]{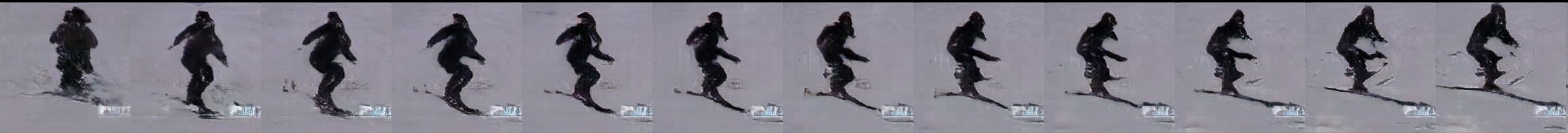} \\
    & \includegraphics[width=.95\textwidth]{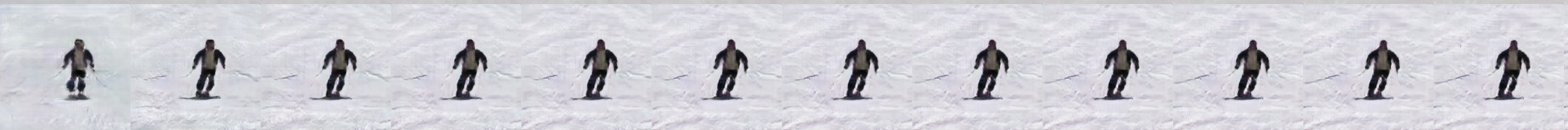} \\[2pt]

    \multirow{3}{*}{\rotatebox{90}{\makebox[0pt][c]{TaiChi}}} &
    \includegraphics[width=.95\textwidth]{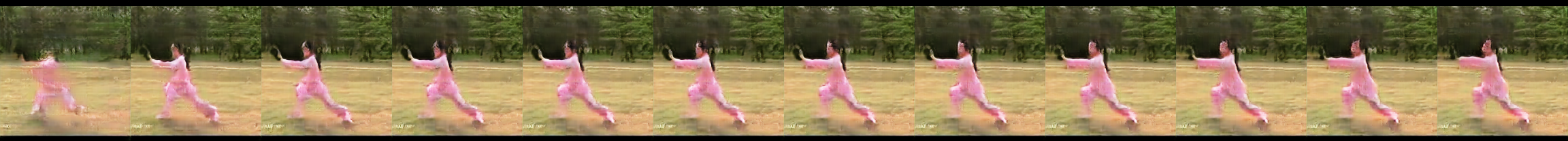} \\
    & \includegraphics[width=.95\textwidth]{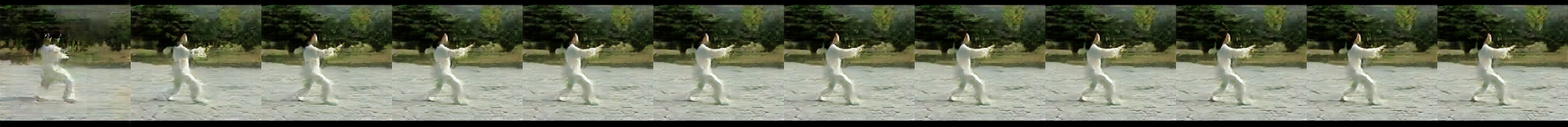} \\
    & \includegraphics[width=.95\textwidth]{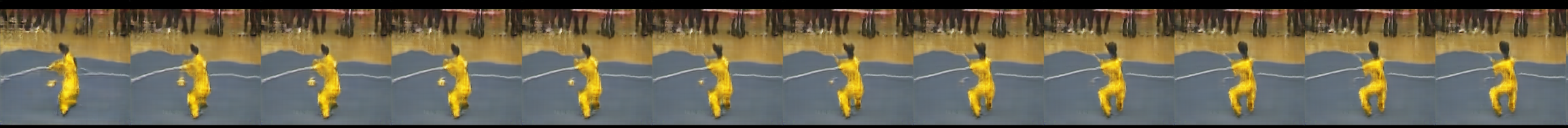} \\
    \end{tabular}    
    \caption{\przemek{An example of video frames generated be TGANv2 model trained on four categories of the UCF-101 dataset. Each row presents twelve generated consecutive frames from the movie belonging to category described in the right part of the Figure. Compare to Fig. \ref{fig:our_generated}, which presents an example of video frames generated by MeVGAN.}}
    \label{fig:our_generated_TG}
\end{figure}

In addition, we compared the model trained this way with the Temporal GAN v2 model. As in our model, we trained TGANv2 separately on selected categories of the UTF-101 set. The example of video frames generated by TGANv2 is presented in Figure~\ref{fig:our_generated_TG}. To evaluate the models, we used the three popular metrics used to evaluate the quality of generative models in the domain of video and image generation: Fréchet Video Distance (FVD)~\cite{unterthiner2019fvd}, Fréchet Inception Distance (FID)~\cite{heusel2017gans}, and Inception Score (IS)~\cite{salimans2016improved}.
\przemek{
FVD and FID are based on the popular Fréchet Distance between $P_R$ and $P_G$ is defined by: $d(P_R, P_G) = min_{X, Y} E|X - Y|^2$. In general case Fréchet Distance is difficult to compute, but when $P_R$ and $P_G$ are multivariate Gaussians the expression has form:
\[
d(P_R, P_G) = |\mu_R - \mu_G|^2 + Tr(\Sigma_R + \Sigma_G - 2(\Sigma_R\Sigma_G)^{\frac{1}{2}})
\]
where $\mu_R$ and $\mu_G$ are the means and $\Sigma_R$ and $\Sigma_G$ are the co-vatiance matrices of $P_R$ and $P_G$. To compute FVD and FID on real world videos and images we pass samples into an Inflated 3D Convnet (I3D) and Inception network respectively to record the activation responses of real and generated samples. Then metrics are computed using the means and covariances obtained from the recorded responses. The IS is also calculated by leveraging the outputs of the Inception network which produces class probabilities. The IS is then calculated by considering the entropy of the class probabilities for each generated sample, representing the diversity, and the average Kullback-Leibler divergence between the marginal class distribution and the overall distribution of classes, representing the quality.  
}
To evaluate the performance of these two models, we generated a total of 4092 video clips, each containing 16 frames, and calculated the Fréchet Video Distance metric for each model. Since other metrics such as Fréchet Inception Distance and Inception Score are designed for image data, we randomly split each video clip into 5 parts and calculated the FID and IS for each split. Finally, we computed the average and standard deviation across all splits for each model. The results of these evaluations are presented in Table~\ref{tab:our_TGAN}.

Our approach utilized a two-stage model learning process that resulted in improved performance compared to previous methods such as TGANv2. In the first stage, we trained our generative model on images to capture the underlying distribution of images' features. In the second stage, we utilized the pre-trained model to train the sequence generation part of the model that creates noise vectors to form video. This two-steps approach allowed the second stage to focus on capturing temporal dependencies in video data, which is critical for generating smooth and continuous video sequences.

The effectiveness of our two-stage approach is reflected in the superior results presented in Table~\ref{tab:our_TGAN}. By using FVD, FID, and IS metrics to evaluate our models, we can see that our approach outperforms the previous state-of-the-art method, TGANv2. Our method is capable of generating video sequences that are not only visually pleasing but also more realistic and diverse. 

\begin{table}[!ht]
    \centering
    \caption{The results of comparison between our method (\our{}) and Temporal GAN v2 (TGANv2) presented in three evaluation metrics: Fr\'echet Video Distanc (FVD), Fr\'echet inception distance (FID), and Inception Score (IS). All calculations was performed on 4092 videos, each containing 16 frames. Since the last two metrics are measured on the images, we randomly divided all frames into five parts and calculated each subset's mean and standard deviation. 
    \przemek{
    A higher IS score is considered 'better,' which stands in contrast to the interpretation of the other metrics (FVD and FID). It's worth noting that the values in these table cells reflect the enhanced efficiency of our approach according to these metrics.}
    }
    \label{tab:our_TGAN}
    \begin{tabular}{@{}l@{\;}l@{\;}rrrr@{}}
    \toprule
        \textbf{Data} & \textbf{Method} & \textbf{FVD} $\downarrow$ & \multicolumn{1}{c}{\textbf{FID $\downarrow$ }} & \multicolumn{1}{c}{\textbf{IS (Fake) $\uparrow$ }} & \multicolumn{1}{c}{\textbf{IS (Real) $\uparrow$}} \\
    \midrule
        \multirow{2}{*}{BalanceBeam} & \our{} & 1000.80 & \bf 74.86($\pm$0.37) & \bf 1.09($\pm$0.0005) & \bf 1.09($\pm$0.0010) \\
        & TGANv2 &  \bf 982.01 & 94.79($\pm$0.23) & 1.08($\pm$0.0005) & 1.07($\pm$0.0008) \\
        \cmidrule(l{3pt}r{3pt}){1-6}
        \multirow{2}{*}{BaseballPitch} & \our{} & \bf 557.42 & \bf 35.68($\pm$0.38) & 1.06($\pm$0.0002) &  \bf 1.06($\pm$0.0001) \\
        & TGANv2 &  608.17 & 77.35($\pm$0.59) & \bf 1.07($\pm$0.0002) & 1.06($\pm$0.0003) \\
        \cmidrule(l{3pt}r{3pt}){1-6}
        \multirow{2}{*}{Skiing} & \our{} & \bf 623.65 & 63.76($\pm$0.41) & \bf 1.09($\pm$0.0005) & \bf 1.09($\pm$0.0007) \\
        & TGANv2 &  660.48 & \bf 61.45($\pm$0.45) &  1.08($\pm$0.0006)  &  1.07($\pm$0.0005) \\
        \cmidrule(l{3pt}r{3pt}){1-6}
        \multirow{2}{*}{TaiChi} & \our{} & \bf 259.39 & \bf 56.19($\pm$0.34) &  1.07($\pm$0.0003) &  1.08($\pm$0.0004) \\
        & TGANv2 &  545.25 & 124.66($\pm$0.65) & \bf 1.11($\pm$0.0008) & \bf 1.10($\pm$0.0007) \\
        \cmidrule(l{3pt}r{3pt}){1-6}
        \multirow{2}{*}{\przemek{Colonoscopy}} & \our{} &  104.60 & \bf 44.66($\pm$0.51) & \bf 1.08($\pm$0.0008) & \bf 1.08($\pm$0.0005) \\
        & TGANv2 & \bf 101.76 & 56.83($\pm$0.61) &  1.06($\pm$0.0002) &  1.08($\pm$0.0012) \\
        
    \bottomrule
    \end{tabular}
\end{table}

\begin{figure}
    \renewcommand{\arraystretch}{0.1}
    \begin{tabular}{ @{}r@{\;}l@{} }

\multirow{5}{*}{\rotatebox{90}{\makebox[0pt][c]{ TGANv2 \qquad \qquad \qquad \qquad }}} 

    & \includegraphics[width=.95\textwidth]{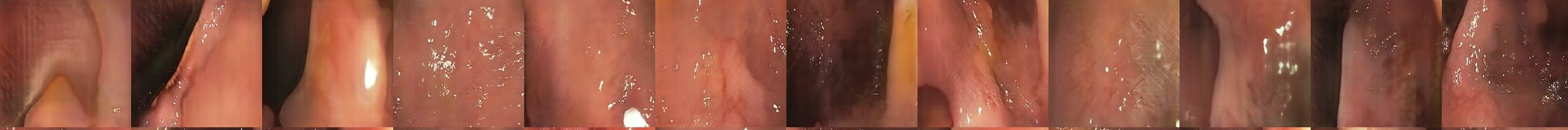}\\
    & \includegraphics[width=.95\textwidth]{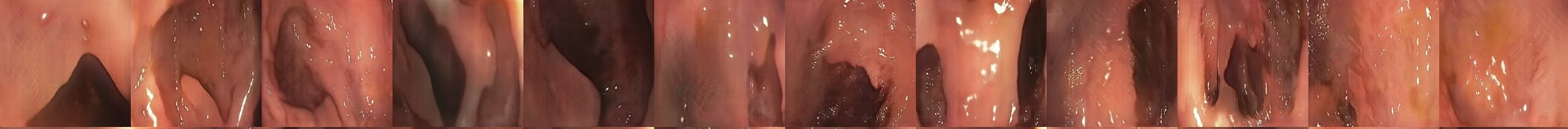}\\
    & \includegraphics[width=.95\textwidth]{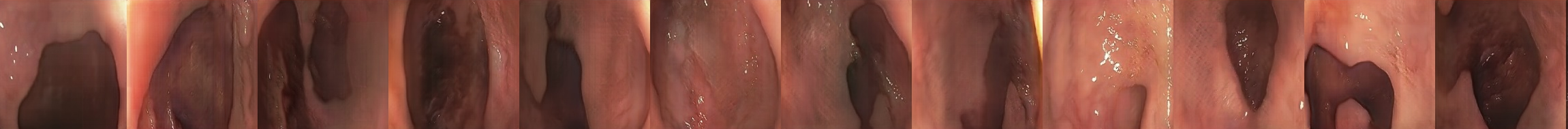}\\
    & \includegraphics[width=.95\textwidth]{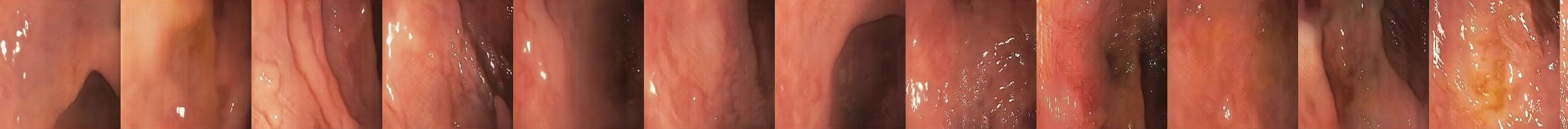}\\
    & \includegraphics[width=.95\textwidth]{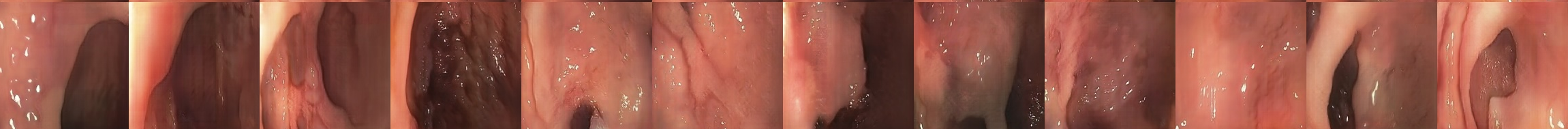}\\ 
\\[1em] 
\multirow{5}{*}{\rotatebox{90}{\makebox[0pt][c]{ \our{} \qquad \qquad \qquad \qquad }}} 
    & \includegraphics[width=.95\textwidth]{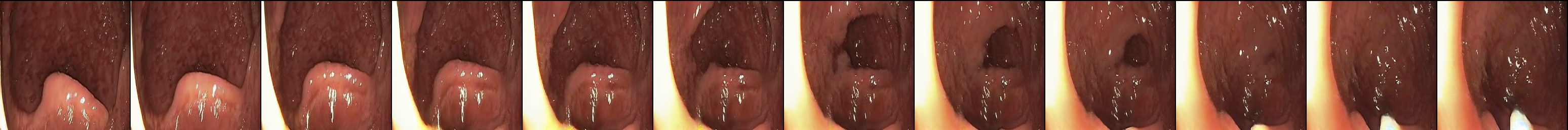} \\
    & \includegraphics[width=.95\textwidth]{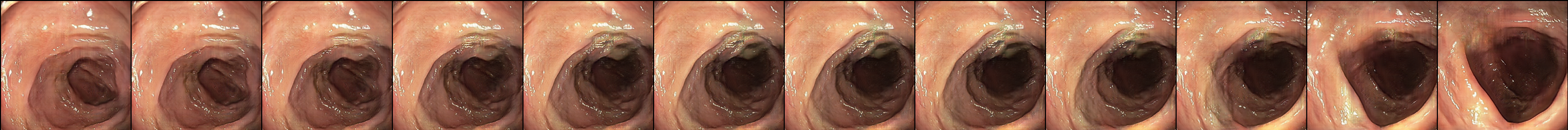} \\
    & \includegraphics[width=.95\textwidth]{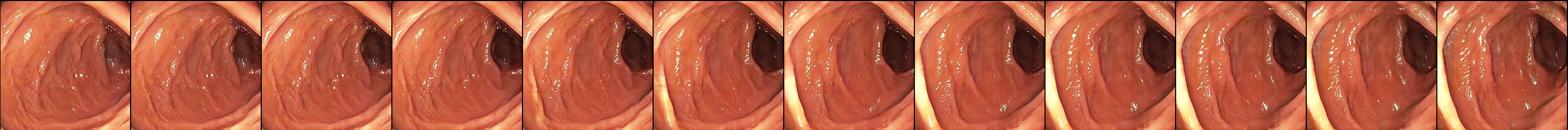} \\
    & \includegraphics[width=.95\textwidth]{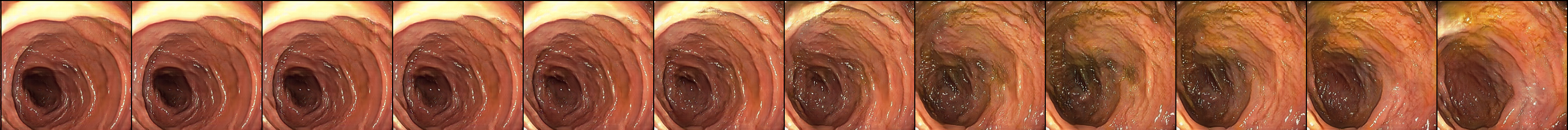} \\
    & \includegraphics[width=.95\textwidth]{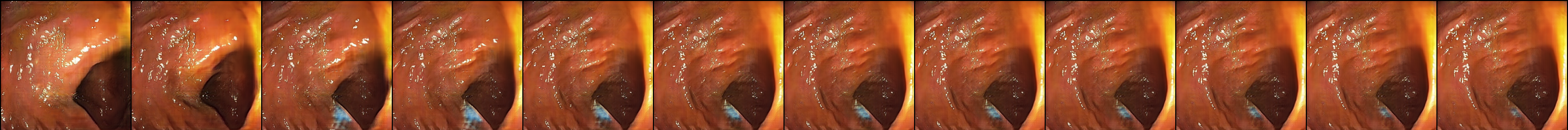} \\

    \end{tabular}  
    \caption{
    \przemek{The examples of frames from five video clips produced by MeVGAN and TGANv2 trained using our colonoscopy data. Each row contains 12 consecutive frame form one clip of $256\times 256$ pixel shape.} }
    \label{fig:colonoscopy_samples}
\end{figure}


\subsection*{Colonoscopy movies}\label{sec.colonoscopy_movies}


Colonoscopy is a medical procedure to examine the large intestine and rectum for abnormalities such as polyps or cancer. This procedure is an important tool for detecting and preventing colon cancer, which is one of the most common types of cancer worldwide.

Simulation-based training is becoming increasingly popular in the medical field, providing a safe and controlled environment for medical professionals to practice and develop their skills. However, using real patient data can be challenging due to ethical and privacy concerns. Generative models can overcome these challenges by generating synthetic medical data that resembles real data while maintaining patient privacy. Furthermore, generative models can create scenarios with specific medical conditions or abnormalities that may be difficult to encounter in real patient data.

\begin{figure}
    \centering
    
    \includegraphics[width=.95\textwidth]{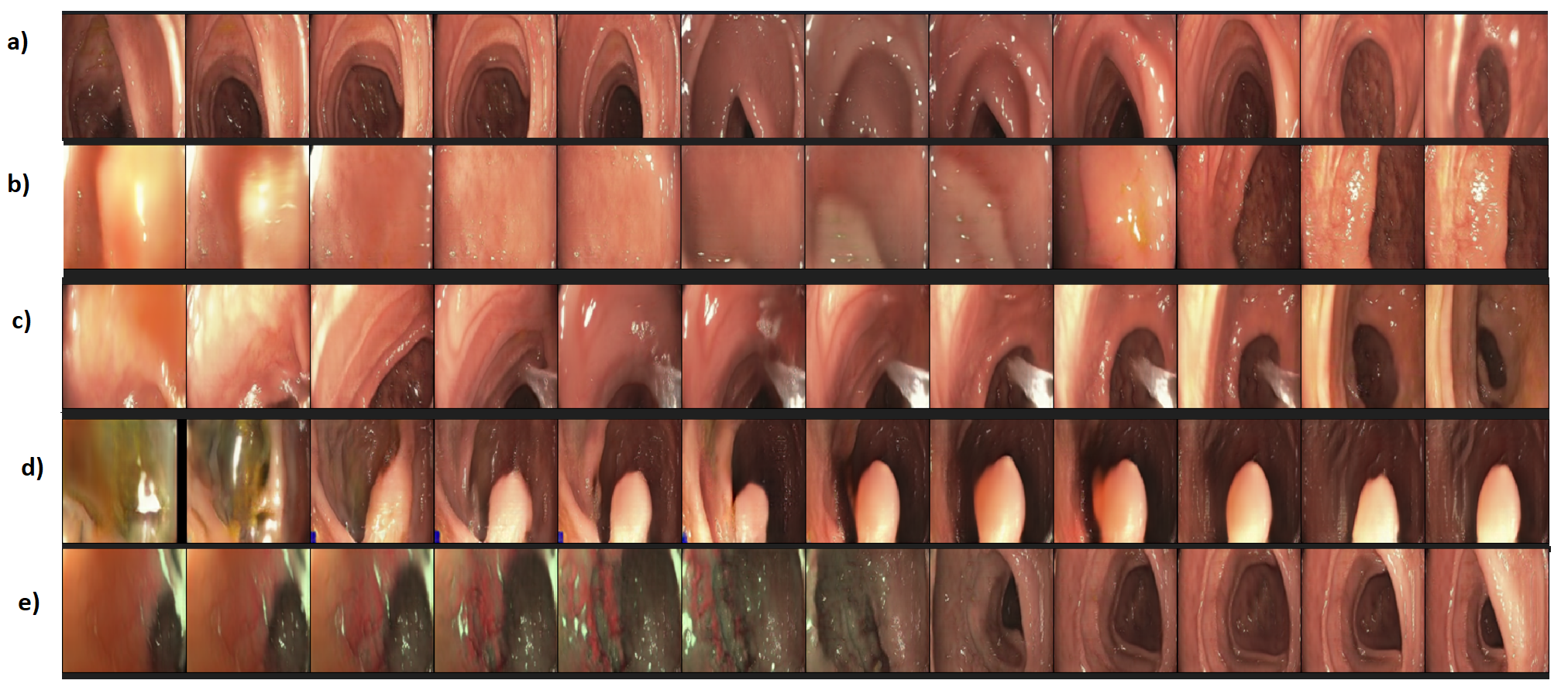}
    \caption{An example of ability of \our{} to generate many  various stages of the colonoscopy procedure e.g. intestine with visible haustration (a), sliding colonoscope on the
intestinal wall (b), rinsing procedure to remove impurity (c), intestine with visible polyp (d) and
ulcerative colitis (e).}
    \label{fig:colonoscopy_samples_op}
\end{figure}

We utilized colonoscopy data for training our generative video model \our{} and achieved promising results. By training our model on a large dataset of real colonoscopy videos, we were able to generate synthetic videos that closely resemble real videos in terms of visual quality and motion patterns, see Figure~\ref{fig:colonoscopy_samples}.
It is important that \our{} can generate many various stages of
the colonoscopy procedure which was presented in Figure~\ref{fig:colonoscopy_samples_op}.

Our approach has several potential applications in the medical field, such as the simulation of colonoscopy procedures for training purposes. Additionally, our model could be used to generate synthetic data with specific conditions or abnormalities that may be challenging to encounter in real colonoscopy videos, thereby aiding in the development and testing of new medical devices and procedures.

\section*{Summary}
This paper presented a novel approach to generating high-quality video sequences using a two-stage model learning process. In the first stage, we trained a ProGAN model on a dataset of images, and in the second stage, we utilized the pre-trained model and additional neural network, to generate sequences of noise vectors, which were used to generate realistic and smooth video sequences. Our approach outperformed the previous state-of-the-art method, TGANv2, in terms of FVD, FID, and IS metrics.

In the case of colonoscopy images, we are able to produce videos with a high level of reality. Moreover, we are able to model many various stages of the colonoscopy procedure e.g. sliding colonoscope on the intestinal wall, rinsing procedure to remove impurity, intestine with visible polyp, and ulcerative
colitis. 

\paragraph{Future works} There are several directions for future research and improvement of our approach. One possible extension is to incorporate more sophisticated techniques for modeling temporal dependencies in video data, such as recurrent neural networks (RNNs) or attention mechanisms. Another potential direction is to explore the use of more complex datasets, such as action recognition datasets, to generate more complex and diverse video sequences. Additionally, it would be valuable to explore ways of further improving the quality and diversity of generated video sequences, such as through the use of adversarial training or fine-tuning the model on specific tasks.

   

\section*{Acknowledgments} 
The project “Design and implementation of a prototype of simulator of colonoscopic examination using virtual and augmented VR/AR reality“ was financed by the Małopolska Centre for Entrepreneurship under contract number RPMP.01.02.01-12-0262/19 from the funds of the Regional Operational Programme for Małopolskie Voivodeship 2014-2020 (first priority axis "knowledge economy").
\nolinenumbers


%
%
%





\end{document}